# Probability of symptoms and critical disease after SARS-CoV-2 infection

Piero Poletti[1,*], Marcello Tirani[2,3,*], Danilo Cereda[2], Filippo Trentini[1], Giorgio Guzzetta[1], Giuliana Sabatino[2], Valentina Marziano[1], Ambra Castrofino[4], Francesca Grosso[4], Gabriele Del Castillo[4], Raffaella Piccarreta[5], ATS Lombardy COVID-19 Task Force[§], Aida Andreassi[2], Alessia Melegaro[5], Maria Gramegna[2], Marco Ajelli[6,#], Stefano Merler[1,#]

[*]corresponding author
[#]senior authors

[1] Bruno Kessler Foundation, Trento, Italy
[2] Directorate General for Health, Lombardy Region, Milano, Italy
[3] Health Protection Agency of Pavia, Pavia, Italy
[4] Department of Biomedical Sciences for Health, University of Milan, Milano, Italy
[5] Bocconi University, Dondena Centre for Research on Social Dynamics and Public Policy, Milan, Italy
[6] Department of Epidemiology and Biostatistics, Indiana University School of Public Health, Bloomington, IN, USA

[§]ATS Lombardy COVID-19 Task Force: Marino Faccini[a], Sabrina Senatore[a], Anna Lamberti[a], Franco Tortorella[b], Silvia Lopiccoli[b], Giancarlo Malchiodi[c], Livia Trezzi[c], Anna Caruana[d], Giovanni Marazza[d], Antonio Piro[e], Luigi Vezzosi[e], Carlo Rossi[e], Annalisa Donadini[f], Jacqueline Frizza[g], Enza Giompapa[g]
[a] Health Protection Agency of Milan, Milan, Italy
[b] Health Protection Agency of Monza-Brianza, Monza, Italy
[c] Health Protection Agency of Bergamo, Bergamo, Italy
[d] Health Protection Agency of Brescia, Brescia, Italy
[e] Health Protection Agency of Valpadana, Mantova, Italy
[f] Health Protection Agency of Insubria, Varese, Italy
[g] Health Protection Agency of Montagna, Sondrio, Italy

## Abstract

We quantified the probability of developing symptoms (respiratory or fever ≥37.5 °C) and critical disease (requiring intensive care or resulting in death) of SARS-CoV-2 positive subjects. 5,484 contacts of SARS-CoV-2 index cases detected in Lombardy, Italy were analyzed, and positive subjects were ascertained via nasal swabs and serological assays. 73.9% of all infected individuals aged less than 60 years did not develop symptoms (95% confidence interval: 71.8-75.9%). The risk of symptoms increased with age. 6.6% of infected subjects older than 60 years had critical disease, with males at significantly higher risk.

## Main

Quantifying the proportion of SARS-CoV-2 infections that do not show recognizable symptoms is still an important missing piece in the puzzle of the ongoing pandemic [1-3]. The role of asymptomatic carriers in transmission is still largely uncertain [4], but episodes of transmission from symptom-free positive subjects have been documented [3,5,6]. Because asymptomatic infections are easily missed by surveillance systems, they can reduce the effectiveness of "test, trace and isolate" strategies in keeping transmission under control and preventing the emergence of local outbreaks [7,8]. Of particular relevance is the topic of transmission of infection by children, given that they are much less likely to experience disease [1,9] and less susceptible to infection than adults and the elderly [10]. Age-specific estimates on the absolute probability of developing symptoms upon infection are still sparse. Similarly, robust estimates of the risk of critical disease (i.e., cases requiring intensive care) upon infection are needed for assessing future scenarios of SARS-CoV-2 spread in terms of their potential healthcare burden [1,11].

In this study, we analyze clinical observations of laboratory confirmed SARS-CoV-2 infections in Lombardy, Italy, to estimate the probability of developing symptoms and experiencing critical disease given SARS-CoV-2 infection, at different ages.

Between February and April 2020, local health authorities (ATS) in Lombardy implemented contact tracing activities to investigate the history of exposure of 64,252 close contacts of 21,410 COVID-19 cases. Cases were ascertained by using a real-time reverse transcription polymerase chain reaction (RT-PCR) assay targeting different genes of SARS-CoV-2 [12,13]. Close contacts of cases were defined as persons living in the same household of a case or who engaged face-to-face with a case within a distance of 2 meters for more than 15 minutes during the exposure period. The exposure period was initially defined as the time interval ranging from 14 days before to 14 days after the date of symptom onset of the index case of the cluster. After March 20, the period was shortened, ranging from 2 days before to 14 days after the symptom onset of the index case [14]. Since the detection of the first locally acquired infection in Lombardy on February 20, 2020 until February 25, all contacts of confirmed COVID-19 cases were tested with RT-PCR, irrespective of clinical symptoms. From February 26 onward, the traced contacts were tested only in case of symptom onset. Overall, 44 PCR tests were performed before February 26; 1,947 afterwards. On April 16, 2020, Lombardy initiated a serological survey aimed at detecting IgG neutralizing antibodies against S1/S2 antigens of SARS-CoV-2 in all contacts of the identified index cases [15,16]. Serological testing was performed using automated LIAISON® SARS-CoV-2 S1/S2 IgG assay (94.4-95.7% sensitivity at 15 days from diagnosis; specificity: 97%-98.5% [15,16]). Serum samples of RT-PCR positive symptomatic subjects were not collected, with the exception of 266 cases. We use preliminary results from the serological survey to integrate the diagnoses of SARS-CoV-2 infection and identify all the infections occurred for a subset of clusters of the traced contacts.

In particular, we selected 3,420 clusters where all contacts were tested against SARS-CoV-2 infection either via nasal swabs during follow-up of contact tracing activities or within the serological survey, for an overall sample of 5,484 close contacts (median age: 50; IQR: 30-61; 56.3% females). Subjects were defined positive to SARS CoV-2 infections if they had at least one laboratory confirmation (either via RT-PCR or serological assay), irrespective of clinical signs. Symptomatic cases were defined as infected subjects showing upper or lower respiratory tract symptoms (e.g. cough, shortness of breath), or fever ≥37.5 °C. Critical cases were defined as patients either admitted to an intensive care unit or deceased with a diagnosis of SARS-CoV-2 infection.

Among the selected 5,484 close contacts (i.e., excluding index cases), 2,824 (51.5%) had been infected (median age: 53; IQR 34-64; 56.8% females). Of these, 1,892 infections (67.0%) were identified by the serological assay only, and 637 (22.6%) by the RT-PCR only, and 295 (10.4%) confirmed by both tests (Table S1).

Of the 2,824 confirmed SARS-CoV-2 infections, 876 (31.0%) were symptomatic. Data stratified by sex, age and testing procedure are displayed in Table 1. The probability of developing symptoms increased with age, ranging from 18.1% (95%CI, 13.9-22.9%) among infected individuals aged less than 20 years to 64.6% (95%CI, 56.6-72.0%) for subjects aged 80 years or more. Positive individuals not showing fever ≥37.5°C or respiratory symptoms represented 73.9% (95%CI, 71.8-75.9%) of all infections detected in individuals younger than 60 years of age. The same quantitative trend was observed in an additional analysis where we considered all of the 6,977 infections detected so far by the ongoing analysis of the contact tracing records (15,836 subjects tested over a total of 64,252 contacts), thus including also clusters for which test results are not available for all individuals (Figure S1A).

75 (2.7%) cases were critical (Tab. 1). Critical care was required for 0.54% of infections occurred under 60 years of age (95%CI: 0. 26-1.00%), in striking contrast with the corresponding proportion among cases above 60 years (6.6%, 95%CI 5.1-8.3%). Male sex was associated with a higher risk of critical disease. The proportion of symptomatic and critical cases among infected males and females are shown in Figure 1.

To test for possible cluster specific effects, we explored the relative risk (RR) of developing symptoms and critical disease in five 20-years-age groups using a generalized linear mixed-effects model (GLMM with logit link and cluster-specific random effects). The model included as covariates the individual's age group and gender, a binary variable indicating whether the cluster's index case was symptomatic or not, and the number of symptomatic contacts in the cluster. A strong age dependency in the risk of symptomatic and critical infection was found, in agreement with the main analysis (Figure S1B). SARS-CoV-2 positive individuals older than 60 years of age showed a significantly higher risk of developing symptoms compared to all younger age groups, while infections under 20 years of age resulted at a significantly lower risk of developing symptoms compared to older infections (Tukey test: p -value<0.001, Figure S1C). No significant differences between females and males were found in the risk of developing symptoms given the infection. However, females resulted 52.7% less likely to experience critical disease (95%CI 24.4-70.7). Similar results were obtained in a sensitivity analysis where we considered all of the 6,977 infections detected in the contact tracing records.

The contribution of asymptomatic infections to transmission of SARS-CoV-2 is still poorly quantified [2,5,17,18]; similar viral loads were found in symptomatic and asymptomatic cases detected with RT-PCR [12,17] and infections caused by asymptomatic subjects were documented [5]. In our sample, the large majority of infections were detected retrospectively via serological testing highlighting the intrinsic difficulties in the detection of infections during surveillance of cases.

Estimates of the proportion of SARS-CoV-2 asymptomatic infections published so far are highly variable, ranging from 17% to 87% [2,5,17,18], and depend on what symptoms are included in the definition and when they were ascertained. Here, we use the definition of any respiratory symptoms or the presence of fever ≥37.5 °C; cases were followed up for symptoms throughout the study period. The proportion of symptomatic infections found in our data is in line with recent modeling estimates [18] and published medical literature [5,17].

The selected sample in our study consists of individuals identified during contact tracing, which may not necessarily represent the general population. However, this design has the advantage of assessing symptoms during active tracing and monitoring of contacts, thereby reducing the possibility of recall bias for individuals identified as positive through serological surveys. In addition, the age distribution of our sample was fairly similar to that of the Italian population. Nonetheless, it is important to stress that the infection attack rates observed in our sample (i.e., close contacts of

COVID-19 cases) is not representative of the general population of Lombardy, as contacts of COVID-19 cases were exposed to a higher risk of infection than the general population. The imperfect sensitivity and specificity of the assays used to confirm infection [15,16,19] should be kept into account when interpreting the results of this analysis; these parameters also change with the time after infection at which the biological sample is collected. Data describing the accuracy of the IgG assays used in this study can be found in [15,16]. Contacts were tested for IgG serology more than 1 month after exposure to their cluster index case, thus false negatives due to delays in seroconversion are likely negligible. However, the presented data cannot be used to compare RT-PCR and IgG testing performances in detecting SARS-CoV-2, because nasopharyngeal swabs were taken at heterogenous delays after infection and the sensitivity of the RT-PCR depends critically on these delays [20].

Our estimates of the age-specific probability of developing symptoms given infection can help separating the contribution of heterogeneous risks of infection [10] and disease [9] by age to the observed age distribution of cases, shedding new light on the epidemiology of COVID-19. The quantification of the risk of critical disease upon infection can inform estimates from transmission models and help ensure the sustainability of health system in terms of supplies, human resources, equipment and beds required at different intensities of care.

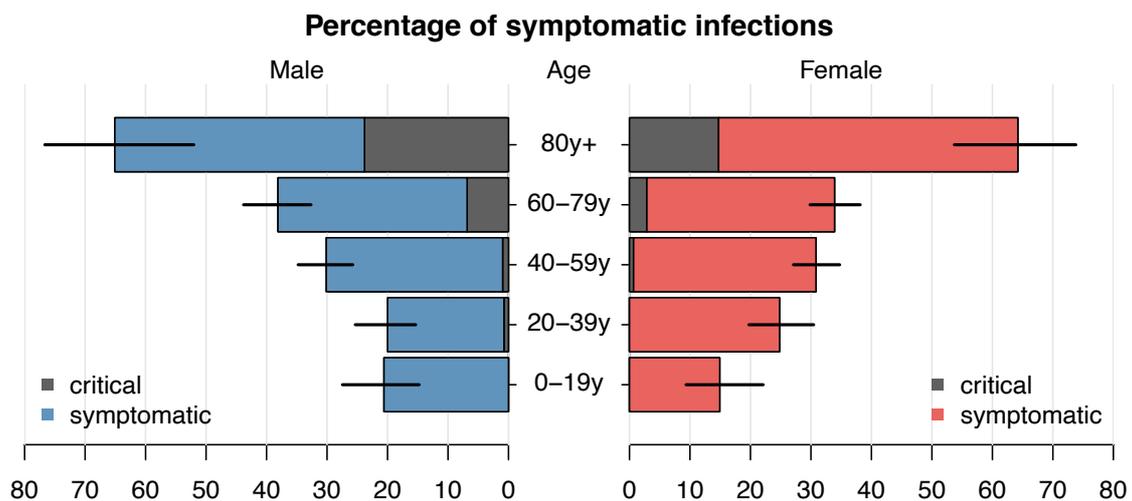

**Figure 1.** Estimated percentage of critical and symptomatic (respiratory or fever ≥37.5 °C) infections in males and females across different age-groups. Horizontal lines represent 95% confidence intervals computed by exact binomial tests.

|  | Subjects | Positive to SARS-CoV-2 infection | Symptomatic infections[#] | | Critical patients | |
|---|---|---|---|---|---|---|
|  |  |  | mean | 95% CI | Mean | 95% CI |
| Gender |  |  |  |  |  |  |
| Male | 2,398 | 1,220 | 371/1,220 (30.41%) | 27.84-33.08% | 42/1,220 (3.44%) | 2.49-4.63% |
| Female | 3,086 | 1,604 | 505/1,604 (31.48%) | 29.22-33.82% | 33/1,604 (2.06%) | 1.42-2.88% |
| Age |  |  |  |  |  |  |
| 0-19y | 692 | 304 | 55/304 (18.09%) | 13.93-22.89% | 0/304 (0%) | 0-1.21% |
| 20-39y | 1,177 | 531 | 119/531 (22.41%) | 18.93-26.2% | 2/531 (0.38%) | 0.05-1.35% |
| 40-59y | 2,015 | 1,002 | 306/1,002 (30.54%) | 27.7-33.49% | 8/1,002 (0.8%) | 0.35-1.57% |
| 60-79y | 1,352 | 829 | 294/829 (35.46%) | 32.2-38.83% | 36/829 (4.34%) | 3.06-5.96% |
| 80+ | 248 | 158 | 102/158 (64.56%) | 56.56-71.99% | 29/158 (18.35%) | 12.65-25.28% |
| Total | 5,484 | 2,824 | 876/2,824 (31.02%) | 29.32-32.76% | 75/2,824 (2.66%) | 2.09-3.32% |

[#] respiratory or fever ≥37.5 °C

**Table 1.** Sample description and estimates by sex, age group and testing procedure.

# Supplementary Figures and Tables

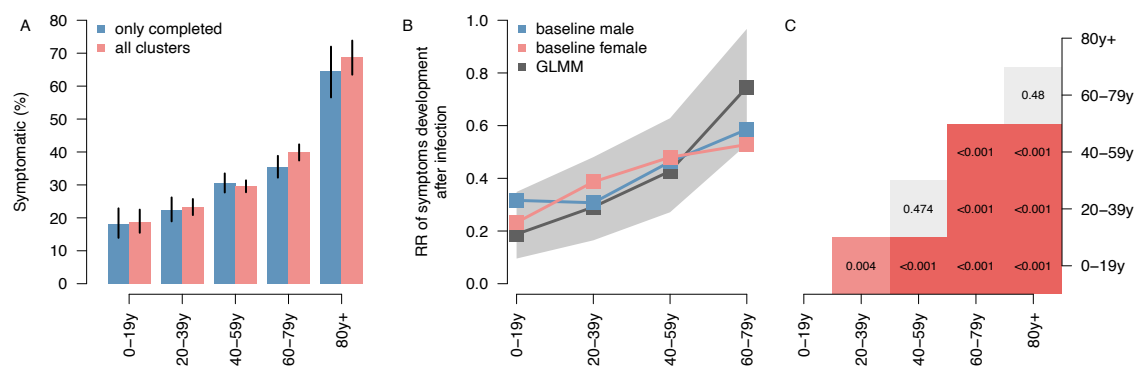

**Figure S1. A)** Probability of developing respiratory symptoms or fever ≥37.5 °C (mean; vertical bars show 95%CI) as estimated by considering only clusters entirely composed by individuals tested either with RT-PCR or serological assay and as estimated by considering all clusters. **B)** Relative risks (RR) of developing symptoms by age group (with the last age group, 80+, as reference) estimated employing a generalized linear mixed-effects model (GLMM) with logit link (shaded area shows 95%CI). The RR increases with the number of symptomatic infections occurred in the cluster (p-value<0.001). **C)** Results of Tukey post-hoc test comparing the RRs across different age classes.

| RT-PCR | Serological assay (IgG) | Total |
|---|---|---|
| Performed | Not performed | 1,364 |
| Positive | - | 632 |
| Not performed | Performed | 3,493 |
| - | Positive | 1,755 |
| Performed | Performed | 627 |
| Positive | Negative | 5 |
| Negative | Positive | 137 |
| Positive | Positive | 295 |

**Table S1.** Description of the sample. From February 26 onward, only symptomatic contacts of COVID-19 confirmed cases were tested with RT-PCR (44 RT-PRC tests were conducted before February 26; 1,947 afterwards).

# Methods

### Sources of data

Soon after the detection of first locally acquired infection in the Lombardy region of Italy on February 20, 2020, national and regional health authorities introduced coordinated interventions to control the epidemic spread. Local health authorities of Lombardy set up a surveillance system to detect local transmission of SARS-CoV-2 by extensive testing, to isolate COVID-19 cases, and to identify and quarantine contacts of positive subjects.

On April 16, the Lombardy Region initiated a large-scale serological screening of subjects quarantined for COVID-19 to evaluate the prevalence of the infection in this group. The data analyzed in this paper were obtained by building a database that combines the information collected during the contact tracing activities conducted between February and April 2020 by the Lombardy healthcare agencies, with the Lombardy linelist of COVID-19 patients and the initial test results obtained through the serological survey currently ongoing in the region. The database provides, among other details, information on the gender and the age of cases and case contacts, the results of PCR tests and serological tests (if any), the outcome of patients, and whether they were admitted to an intensive care unit. Data were collected as part of surveillance activities to control SARS-CoV-2 transmission, so that information on PCR and serological test results consists only of categorical data (i.e., positive, negative, or inconclusive).

The identification and monitoring of close case contacts was performed by regional healthcare agencies, through standardized epidemiological investigations of positive cases (or of their relatives) to determine the history of individuals' exposure. The exposure period was initially defined as the interval time ranging from 14 days before to 14 days after the symptom onset of the index case. After March 20, the time period was reduced to from 2 days before to 14 days after the symptom onset of the index case [14].

According to criteria initially defined by the European Centre for Disease Prevention and Control (ECDC), from February 21 to February 25, suspected COVID-19 cases were identified as:
1. patients with acute respiratory tract infection OR sudden onset of at least one of the following: cough, fever, shortness of breath AND with no other aetiology that fully explains the clinical presentation AND at least one of these other conditions: a history of travel to or residence in China, OR patient is a health care worker who has been working in an environment where severe acute respiratory infections of unknown etiology are being cared for;
2. OR patients with any acute respiratory illness AND at least one of these other conditions: having been in close contact with a confirmed or probable COVID-19 case in the last 14 days prior to onset of symptoms, OR having visited or worked in a live animal market in Wuhan, Hubei Province, China in the last 14 days prior to onset of symptoms, OR having worked or attended a health care facility in the last 14 days prior to onset of symptoms where patients with hospital- associated COVID-19 have been reported.

Probable cases were defined as suspect cases for whom testing for virus causing COVID-19 was positive with a specific real-time RT PCR assay detecting the SARS-CoV-2 virus responsible for the COVID-19 (according to the test results reported by the laboratory) OR for whom testing inconclusive. At any time, confirmed cases were defined as persons with laboratory confirmation of virus causing of SARS-CoV-2 infection, irrespective of clinical signs and symptoms. The Lombardy linelist of patients used to build the database analyzed here consists in all confirmed COVID-19 cases detected until June 8, 2020.

A close case contact was defined as a person living in the same household as a COVID-19 confirmed case, a person having had face-to-face interaction with a COVID-19 confirmed case within 2 meters and for more than 15 minutes; a person who was in a closed environment (e.g. classroom, meeting room, hospital waiting room) with a COVID-19 confirmed case at a distance of less than 2 meters for more than 15 minutes; a healthcare worker or other person providing direct care for a COVID-19 confirmed case, or laboratory workers handling specimens from a COVID-19 confirmed case without recommended personal protective equipment (PPE) or with a possible breach of PPE; a contact in an aircraft sitting within two seats (in any direction) of a COVID-19 confirmed case, travel companions or persons providing care, and crew members serving in the section of the aircraft where the index case was seated (passengers seated in the entire section or all passengers on the aircraft were considered close contacts of a confirmed case when severity of symptoms or movement of the case indicate more extensive exposure). Close case contacts were initially considered as contacts occurred between 14 days before and 14 days after the onset of symptoms in the case under consideration. After March 20, close case contacts were defined as contacts occurred between 2 days before and 14 days after the case symptom onset [14]. Clusters of contacts were defined as the set contacts identified by the contact tracing triggered by a positive index case. Contact data used to build the analyzed database consist of records collected between February 21 and April 16, 2020. Contact data collected after April 16, 2020 were excluded to avoid biases caused by reporting delays, delays between exposure and symptom onset, and in seroconversion of positive individuals.

According to the WHO recommendations, nasal swabs (UTM viral transport ®, Copan Italia S.p.a) from all suspected cases were tested with at least two real-time RT PCR assays targeting different genes (E and RdRp) of SARS-CoV-2 [13]. In addition, a novel quantitative real-time RT PCR targeting an additional SARS-CoV-2 gene (M) was developed (details provided upon request). From February 21 to February 25, all suspected cases and asymptomatic contacts were tested. From February 26 onward, testing was applied only to symptomatic patients. From March 20, positivity to the nasal swab was also granted for tests that sought a single gene. Inconclusive swabs were repeated to reach the diagnosis.

Serological screening of subjects quarantined for COVID-19 included both symptomatic and asymptomatic case contacts identified through an epidemiological investigation without history of a swab for SARS-CoV-2. The test used to detect SARS-CoV-2 IgG antibodies is the LIAISON® SARS-CoV-2 test (DiaSorin). The LIAISON® SARS-CoV-2 test employs magnetic beads coated with S1 & S2 antigens [15,16]. The antigens used in the tests are expressed in human cells to achieve proper folding, oligomer formation, and glycosylation, providing material similar to the native spikes. This strategy ensures that the antigen-antibody complex forms with the required specificity. The S1 and S2 proteins are both targets to neutralizing antibodies. The test provides the detection of IgG antibodies against S1/S2 antigens of SARS-CoV-2 and the detection of neutralizing antibodies with 97,8% negative agreement and 94.4% positive agreement to Plaque Reduction Neutralization Test (PRNT). The comparison to PRNT was evaluated by testing 304 samples collected during the outbreak from subjects whose PRNT result was available. Performance analyses validating the accuracy of IgG serological tests used can be found in [15.16]. A negative result (<12 AU/mL) indicates the absence or a very low level of IgG antibodies directed against the virus, this occurs in the absence of infection or during the incubation period or in the early stages of the disease. An inconclusive result (12-15 AU/mL) can be interpreted as both a false positive or a false negative and suggests repeating the exam after a week. A positive result (>15 AU/mL) indicates the presence of IgG antibodies and must be interpreted in association with the clinical outcomes and the possible search for the viral genome on the nasopharyngeal swab. Clusters with case contacts with inconclusive results for both PCR and IgG tests were excluded by the proposed analysis. The performed analysis is based on all serological test results obtained within June 15, 2020 from serum samples collected before May 24, 2020.

## Statistical analysis

As of the time of writing this manuscript (June 22, 2020), the serological survey aiming at completing the ascertainment of SARS-CoV-2 infections among all close contacts of COVID-19 confirmed cases is still ongoing. Therefore, only a fraction of the asymptomatic infections has been detected by the serosurvey thus far; whereas, the majority of symptomatic infections had already been confirmed by RT-PCR. As such, to avoid a biased sample where all symptomatic individuals are included, while only a fraction of asymptomatic infections are considered, we analyzed only those clusters where all contacts were tested (either via RT-PCR or serological assay). Nonetheless, as a sensitivity analysis, we analyze also all contacts with at least a laboratory result.

Contact tracing data were combined with test and clinical outcomes of close contacts associated with each index case and used to estimate the absolute probability of relative risk of infection and of developing symptoms by age. We categorized contacts in five 20-years-age groups (0-19 years, 20-39 years, 40-59 years, 60-79 years, 80+ years). Probabilities stratified by age and gender were defined as the proportion of symptomatic infected individuals and critical patients among the total number of infected individuals. Exact binomial test was applied to compute confidence intervals for different considered strata.

We applied a logistic regression model where the clinical outcome of positive close contacts is considered as the response variable and using the following covariates:
1) age group (of the contact);
2) gender (of the contact);
3) a binary variable defining whether the index case in the cluster was symptomatic or not;
4) the number of symptomatic individuals in the cluster.

Risk ratios of experiencing symptoms were computed given the covariates. Resulting means were compared by Tukey post hoc test. Specifically, to account for the clustering in the binary data, we applied a generalized linear mixed-effects model (GLMM) with logit link specified as follows:

$$m(\mu_{ic}) = \alpha + \beta_1 A_{ci} + \beta_2 G_{ci} + \beta_3 P_{ci} + \beta_4 N_{ci} + u_c$$

where $m$ is the logit link function, $\alpha$ is the intercept, $A_{ic}, G_{ic}, P_{ci}$ and $N_{ci}$ denote respectively the fixed effects of age group an individual $i$ belongs to, the gender of the contact, a binary variable indicating whether the exposure took place in a cluster with a symptomatic index case, and the number of symptomatic infections in the cluster; $u_c$ is the cluster-specific random effects and $\mu_{ic} = E(Y_{ci}|u_c)$ is the mean of the response variable $Y_{ci}$ for a given value of the random effects. The statistical analysis was performed with R (version 3.6).

## Ethics approval

Data collection and analysis were part of outbreak investigation during a public health emergency. Processing of COVID-19 data is necessary for reasons of public interest in the area of public health, such as protecting against serious cross-border threats to health or ensuring high standards of quality and safety of health care, and therefore exempted from institutional review board approval (Regulation EU 2016/679 GDPR).

# Acknowledgements

PP, FT, GG, VM, and SM acknowledge funding from the European Commission H2020 project MOOD and from the VRT Foundation Trento project "Epidemiologia e transmissione di COVID-19 in Trentino"

## Competing interests

The authors declare no competing interests.

## Contributions

PP,MA,SM conceived and designed the study. PP performed the analysis. MT,DC,GS,AC,FG,GDC,AA,MG collected data. PP,MT,DC,FT collated linked clinical–epidemiologic data. MT,DC verified all data. PP,GG,MA wrote the first draft. All authors contributed to data interpretation, critical revision of the manuscript and approved the final version of the manuscript.